# Shape-Adaptive Motion Estimation Algorithm for MPEG-4 Video Coding

F. BENBOUBKER, F. ABDI and A. AHAITOUF

Signals Systems & Components Laboratory (SSC)
Faculty of Sciences and Techniques, FES,
BP 2202, FEZ; MOROCCO.

**Abstract**
This paper presents a gradient based motion estimation algorithm based on shape-motion prediction, which takes advantage of the correlation between neighboring Binary Alpha Blocks (BABs), to match with the Mpeg-4 shape coding case and speed up the estimation process. The PSNR and computation time achieved by the proposed algorithm seem to be better than those obtained by most popular motion estimation techniques.
***Keywords:*** *Motion Estimation algorithm, BAB, Shape Coding, MPEG4.*

## 1. Introduction

Motion estimation and compensation is a key component for high quality video compression, which is characterized by its high computation complexity and memory requirements. However, Motion estimation is considered as the most time-consuming stage in MPEG processing [1] (up to 90% of the total execution time [2]). Therefore, to achieve performances desired for real time applications, it's imperative to think about hardware architecture and use a motion estimation algorithm which reduces computation complexity.

The best performances, in term of PSNR, are achieved by exhaustive search (ES) ME algorithms, since they examine all possible motion vectors, however, their implementation increase the computation time and slow down the compression process [3]. Fast search algorithm, such as 2-D log search scheme [4], the Three step search (TSS) [5], the Four Step Search (FSS) [6] and Diamond Search (DS)[7] have been proposed, all of them try to achieve the same PSNR as the ES by considering only the most probable motion vectors.

In fact, many researchers have focused on ME algorithms especially based on texture coding. However, one of the most important concepts introduced by the Mpeg-4 visual standard is the use of video object (VO) as an entity the user can access and manipulate. The instance of a VO at a particular point of time is called video object plane (VOP) [8]. To support coding of arbitrary-shaped objects, each position in the picture is associated to a Binary Alpha Blocks (BAB); and thus macro-blocks of the image are classed as: opaque (fully 'inside' the VOP), transparent (not part of the VOP) or on the boundary of the VOP. Therefore, in MPEG-4 video coding, ME of shape is also imperative for real-time VOP-based encoding.

Several papers have proposed software implementation methods for shape coders [9], [10] where shape information is used to reduce search point per macro-block and only valid predecessors are evaluated [10] for boundary macro-blocks.

Since hardware implementation is usually better to achieve the complexity suitable for real-time applications, we propose in this document a gradient based algorithm where ME for shape coding is combined with ME for texture, which we will use for a hardware implementation of an MPEG4 encoder IP to accelerate convergence process. The algorithm uses shape ME for boundaries macro-blocks and textures ME for opaque macro-blocks.

To check its performances, we have implemented and tested the proposed algorithm with many test video sequences. Results show that the algorithm presents a good PSNR result with a net decreasing in the number of iterations and computation time. The next section presents background information about video coding and motion estimation, the main idea of the proposed algorithm is described in the section 3 and the evaluation of obtained results is presented in section 4.

## 2. Motion Estimation in Video Coding

For video compression case, the goal is to remove the redundancy in images and reduce the amount of bits required to represent the video sequence. In addition to the discrete cosine transform (DCT) and the quantization block used to remove spatial redundancy, a typical MPEG encoder utilizes a motion estimation (ME) and compensation system to remove temporal redundancy between successive frames of the treated video.

In block-based video coding standards such as Mpeg-4, the first video encoding stage performs motion estimation and compensation for each frame of the video sequence. In this step, we compare the content of the current and previous images and encode only displaced difference blocks, with motion vectors, instead of encoding all original blocks. Conventional algorithms generally use Matching-based or Gradient-based techniques to compute motion vectors.

**Matching-based techniques:** in these approaches, true motion vectors can be determined based on the differences of pixel intensities. The best matching is obtained for smallest differences between pixel intensities of the current and reference frames.

**Gradient-Based techniques:** in these approaches, based on the "intensity conservation over time assumption", the spatiotemporal derivatives of pixel intensities is measured to determine true motion vectors. The total derivative of the image intensity function (I) should be zero every time and for each position in the image:

$$\frac{\partial I}{\partial x}\frac{dx}{dt} + \frac{\partial I}{\partial y}\frac{dy}{dt} + \frac{\partial I}{\partial t} = 0 \qquad (1)$$

In the search process, the problem is to find the motion vector MV for the current block $B_\tau(y,x)$ at time instance $\tau$, so that the error SAD (sum of absolute differences) between the block $B_\tau$ and the matching block $C_\tau$ at time instance $\tau$ is minimized.

## 3. Proposed Algorithm

For commonly used motion estimation algorithms, there is no limit on the number of steps that the search algorithm can take. Therefore we thought about exploiting the optical-flow principle and use a recursive motion estimation which is a less complex method to compute dense displacement fields [10]. The proposed algorithm can be divided into two main steps as shown in Fig. 2: the first step is a Block recursive search, where four candidate vectors (three spatial and one temporal) are evaluated for the actual block by recursive block matching. The second step is a Pixel recursive search, where the chosen vector is adjusted by a gradient based method to find the best approximation.

Fig. 1: Procedure in the proposed motion estimation algorithm

```
h_max = Nbpixel / Block_size
v_max = Nbline / Block_size
for v = 1 : v_max
  for h = 1 : h_max
    X=(h,v)%% current position
   %%%%%%%%%% Block-Recursive Search%%%%%%%%%%
    Type = Block_Type[h,v]%% to define the type of the current
                            macro-block depends on its BAB %%
    if type == Transparent %% no motion vector is computed
       MV(h,v)= (0,0);
    else
       %% temporal candidate %%
       T(h,v)=MV(X); %% motion vector computed for the
                                       previous image %%

     %% computation of neighboring blocks coordinates %%
       A(h,v)= MV(h-1,v); %%motion vector computed for the
                                      block A %%
       B(h,v)= MV(h,v-1); %%motion vector computed for the
                                      block B%%
       C(h,v)= MV(h+1,v-1);%%motion vector computed for the
                                      block C%%

     if type == Boundary
  %% distortion will be measured based on alpha masks of the
     current image(BAB) and the previous one (BAB_P) %%
       SAD_A = SAD[BAB_P(X), BAB(X+A)];
       SAD_B = SAD[BAB_P(X), BAB(X+B)];
       SAD_C = SAD[BAB_P(X), BAB(X+C)];
       SAD_X = SAD[BAB_P(X), BAB(X+T)];
  %% computation of the minimum distortion
       [position,sad_min] = Min[SAD_A,SAD_A,SAD_A,SAD_X]

     else %%    type == Opaque
  %% distortion will be measured based on video data of the
     current image(VID_C) and the previous one (VID_P) %%
       SAD_A = SAD[VID_P(X), VID_C(X+A)];
       SAD_B = SAD[VID_P(X), VID_C(X+B)];
       SAD_C = SAD[VID_P(X), VID_C(X+C)];
       SAD_X = SAD[VID_P(X), VID_C(X+T)];
  %%computation of the minimum distortion
       [position,sad_min] = Min[SAD_A,SAD_A,SAD_A,SAD_X]
    end
 end
        switch position
            case 1,
                MV(X)=A(h,v);
            case 2,
                MV(X)=B(h,v);
            case 3,
                MV(X)=C(h,v);
            case 4,
                MV(X)=T(h,v);
        end
        %%%%%%%%% Pel-Recursive Search%%%%%%%%%%
        %% update of the computed vector based
        %% on the spatio-temporal gradient
        %%%%%%%%%%%%%%%%%%%%%%%%%%%%%%%%%%%%%%%
        MV(X)=DPD(MV(X));
  end
end
```

### 3.1 Block Recursive Search (BRS)

Each video frame is scanned from the top left to the bottom right at sparse block grid to evaluate candidate vectors depending on the macro-block type (opaque, transparent or boundary).

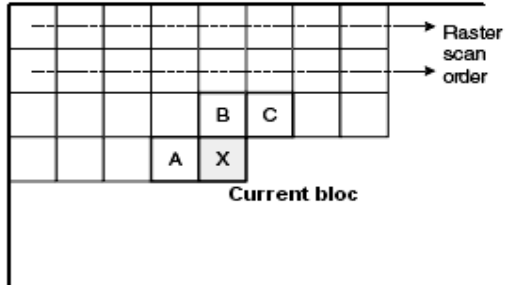

Fig. 3: Neighboring blocks for motion vector selection.

As shown in Fig. 3, the motion vector is selected between the three motion vectors of the neighboring blocks (A, B and C) and the temporal motion vector of the current block (X). For each vector we compute the motion compensation error by computing the SAD of the current block Bτ(y,x) and the predicted one Cτ. The best selection corresponds to the motion vector which minimizes the SAD.

Since transparent macro-blocks are not part of the video objects, no vectors are evaluated for these macro-blocks. For boundary macro-blocks, since they mainly contain shape information, a shape ME is processed. The three spatial vectors are evaluated by referring to shape and motion vectors are evaluated by computing the compensation error based on the BABs around the processed macro-block [9]. The temporal candidate vector is evaluated by referring to texture.

For opaque macro-blocks, a texture ME is processed; motion information is calculated by referring to texture around the processed macro-block. The four candidate vectors are evaluated by computing the compensation error based on texture information.

3.2 Pixel Recursive Search (PRS)

This stage is used to refine ME process; the principle is to update the value of the selected vector in respect to a gradient based technique.

The displacement vector $\vec{d}$ at the current position is obtained as follow [11]:

$$\vec{d}(x,y) = \vec{d}_i - \varepsilon \cdot DPD(\vec{d}_i, x, y) \cdot \frac{\overrightarrow{grad}\left[f(x,y)\right]}{\left\|\overrightarrow{grad}\left[f(x,y)\right]\right\|^2} \quad (2)$$

where ε is the so-called convergence factor.

The displaced pixel difference (DPD) is computed iteratively till its minimum value is reached, based on the selected vector "$d_i$" which corresponds to the minimum value obtained from the BRS.

By replacing the gradient function in the equation (2) by its approximation, the displacement vector equation will be:

$$\vec{d}(x,y) = \vec{d}_i - \varepsilon \cdot DPD(\vec{d}_i, x, y) \cdot [u_x, u_y]^T \quad (3)$$

With

$$u_x = \begin{cases} 0, & \text{if } \frac{\delta f(x,y)}{\delta x} < \Theta \\ \left[\frac{\delta f(x,y)}{\delta x}\right]^{-1}, & \text{else} \end{cases} \quad (4)$$

And

$$\frac{\delta f(x,y)}{\delta x} \approx \frac{f(x+1,y) - f(x-1,y)}{2} \quad (5)$$

Where f(x) is the pixel's gray level, at the location given by the x position and Θ is a threshold value which decreases the sensitivity of pixel recursion to noise; it is usually set to a value of two or three[10].

The Corresponding equation for $u_y$ is obtained by exchanging the index.

In Mpeg-4 visual standard the default block size for motion compensation is 16×16, to improve compression efficiency the standard support four motion vectors per macro-block. Therefore in recursive search we will work with 8×8 blocks.

The PRS compare the DPD in the current position, pointed by the selected vector in the BRS, with the PDP in the others positions which are obtained by shifting the predicted block with an update vector MV in the eight directions (Fig.4).

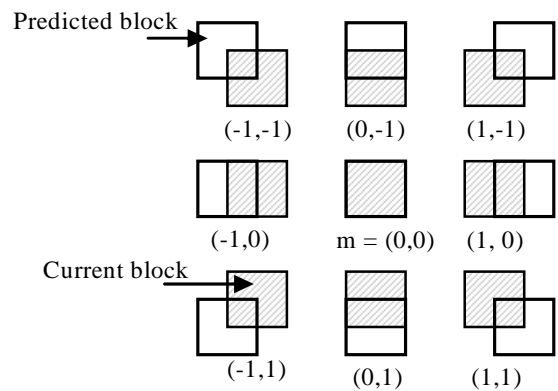

Fig. 4: candidates in the pixel recursive search

The final motion vector will be computed for the position with the smallest DPD.

## 4. Experimental Results

To check its performances, the proposed algorithm is evaluated with four MPEG-4 test video sequences of QCIF format (176x144); Caltrain, Weather, foreman and Carphone. The Caltrain video sequence contains several moving objects on textured background; Weather and Carphone are low-motion video clips, while Foreman contains some quick motion scenes.

All tested video scenes are used to generate the frame-by-frame motion vectors, with two frames distance between current frame and reference frame.

Fig. 5 shows a reference frame from "caltrain" sequence and the corresponding reconstructed image with motion compensation. Motion estimation is carried out by using the proposed Algorithm with a 16×16 block size and half-sample accuracy producing the set of vectors shown on the same figure.

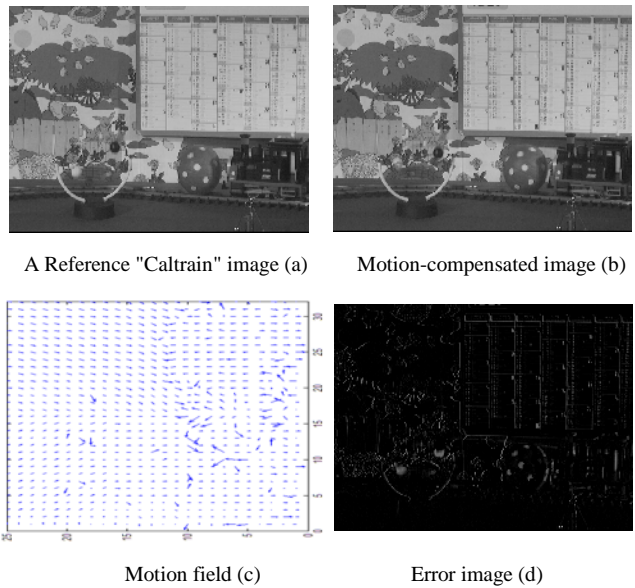

A Reference "Caltrain" image (a)    Motion-compensated image (b)

Motion field (c)    Error image (d)

Fig. 5 : Example of a "Caltrain" frame with its corresponding predicted image, motion vectors and error image

The comparison between the reference and the reconstructed images as well as the residual image (Fig. 5-d) can inform about the proposed algorithm efficiency and performances. We see that there is no significant visual difference between the reference and the reconstructed images, and that the difference image doesn't contain a significant energy, that means the estimation performed by the algorithm is good.

The same video sequence is used to compare the performance of the proposed algorithm with the performance of motion estimation techniques, presented above, whose are widely accepted by the video compressing community and have been used in the implementation of various standards. Motion-compensated images, created from motion vectors, are compared to the reference frame by computing the Peak-Signal-to-Noise-Ratio (PSNR).

In the ES case, since we compare the current block with all blocks in the search window, it corresponds to the highest PSNR values. Fast algorithms attempt to achieve the same PSNR as in the ES wit h minimum computations.

Figures Fig.6 and Fig.7 show respectively PSNR (in dB) results and computation time obtained for "caltrain" sequence. Experimental results demonstrate that the proposed algorithm (P.A) have a comparable PSNR as the ES algorithm and achieves consistent improvement in PSNR over the TSS algorithm which has been widely accepted as one of the best ME for low bit rate real time video applications [12,13].

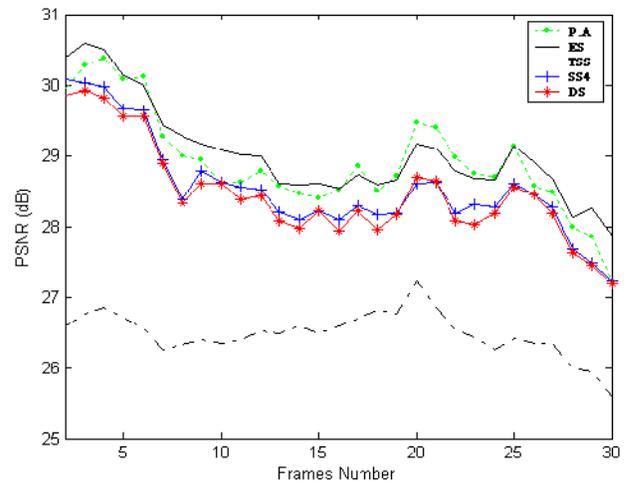

Fig. 6: PSNR Results for "Caltrain" Sequence

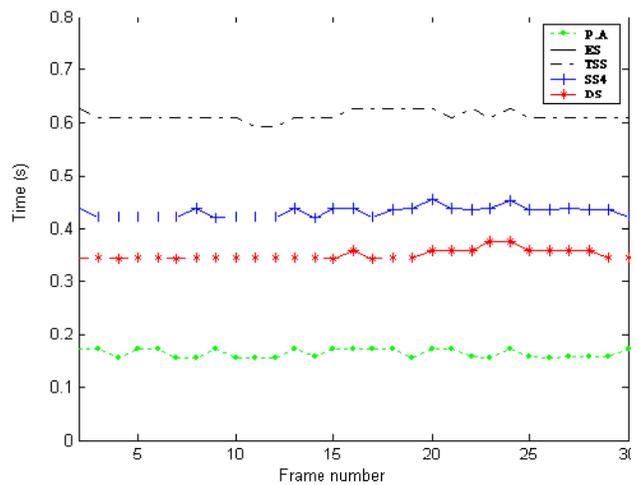

Fig. 7: Computation time for Caltrain sequence

Although PSNR results obtained for DS and FSS are relatively the same as the P.A, they present a higher computational time

To compare the performance of tested algorithms by considering both estimation quality and computation time, we use an indicator as a "figure of merit" which takes into account, for each test sequence, the average value of the obtained PSNR (in linear scale) , the number of macro blocks used in the comparison and the search window. The indicator is defined as:

$$I = \frac{PSNR(2S+1)^2}{C} \qquad (6)$$

Where C is the average number of compared blocks and S is the size of the search window. $(2S+1)^2$ represent the total number of blocks in the search area S. The optimum value of the Indicator will match to a maximum of both PSNR and the ratio of the total number of the blocks in the area search to the number of the compared blocks. Good performance of the algorithm corresponds to high values of the indicator.

Table 1 shows the average of search points per macro-block for tested sequences obtained for a search window size of 7x7. While the ES test around 205 search points per macro-block, the other tested ME algorithms accomplish a good performances with a higher speed-up ratio. For all tested algorithms, even if the number of comparison required per macro-block is clearly reduced by reference to ES, an average of 15 search points for DS, the proposed algorithm presents the best computation time and drop down the number of comparison required per macro-block to an average of 6.5 search points.

Table 1: Average of search points per macro-block for tested ME algorithms

| Sequence | Caltrain | Foreman | Weather | Carphone |
|---|---|---|---|---|
| ES | 204.2828 | 204.2828 | 204.2828 | 204.2828 |
| TSS | 24.3838 | 23.2916 | 23.1343 | 22.5824 |
| 4SS | 20.4460 | 18.4625 | 15.7979 | 17.5911 |
| DS | 15.6392 | 16.1977 | 12.2388 | 15.2116 |
| PA | **6.5125** | **6.4867** | **6.4780** | **6.4286** |

Table 2 shows average values of PSNR obtained for each video sequence. Results demonstrate that the proposed algorithm (P.A) have nearly same results as the ES algorithm for the four sequences and achieves consistent improvement in PSNR over the TSS algorithm.

Table 2: Average values of PSNR (dB) obtained for tested ME algorithms

| Sequence | Caltrain | Foreman | Weather | Carphone |
|---|---|---|---|---|
| ES | 30,54 | 38,52 | 40,68 | 38,55 |
| TSS | 27,53 | 38,02 | 40,28 | 37,93 |
| 4SS | 29,06 | 38,11 | 40,37 | 38,06 |
| DS | 29,56 | 38,35 | 40,63 | 38,43 |
| PA | **29,66** | **38,43** | **40,65** | **38,48** |

Although PSNR results obtained for DS and FSS are relatively the same as the P.A for video scenes with low motions (Weather, Carphone), its PSNR performances are better than them for video scenes which present a quick or complex motions (Foreman, Caltrain) even it checks less search points.

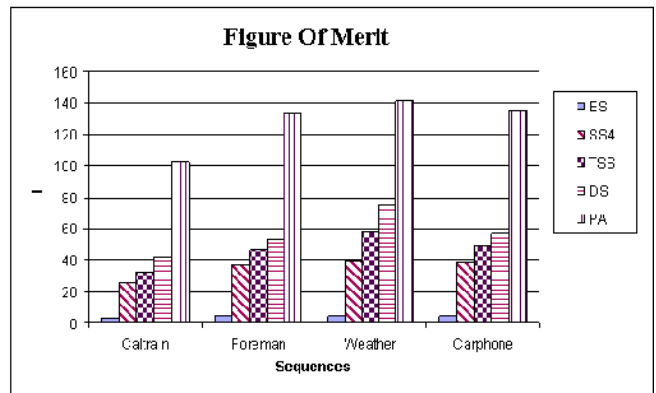

Fig. 8: Performance indicator (Figure of merit)

Fig. 8 presents the indicator values obtained for each test sequence. The figure shows that the proposed algorithm achieves the best compromise between computation time and PSNR results for all tested scenes.

Furthermore, while the processing time may depend on frames and video sequences for most of the algorithms, results show that the P.A keeps the same computation time for all frames and video sequences. Which mean that the ME processing will not depend on the treated video data.

## 4. Conclusions

In this work we have proposed a new efficient algorithm for motion estimation based on the spatio-temporal gradient which uses block and pixel recursive search. The algorithm is based on the shape motion estimation and takes advantage of the texture-shape correlation.
Simulations show that the proposed algorithm reduces the number of comparisons and computation time required for motion estimation process with negligible quality degradation. When compared to commonly used algorithms, the proposed algorithm gives the best PSNR results, very close to those obtained by the ES algorithm, with a number of compared blocks neatly reduced, regardless the kind of treated video. Also, for real time applications, one can take advantage of the invariable computations of the algorithm to control and reduce processing delay.

**Fahd Benboubker:** Masters Degree in Microelectronics & Telecommunication systems (2003-2005), Bachelors Degree in Telecommunication (1999-2003); R&D Engineer at Lead Tech Design, currently Project manager at MarsaMaroc; he is currently working toward the Ph.D. degree in the Department of Electrical Engineering. His major research interests include VLSI architecture design and algorithms for Image, Audio and video processing, reconfigurable computing for multimedia systems. He is a member of the LSSC laboratory.

**Farid ABDI:** received the Ph.D. degrees in Physics from the Metz University in France 1992. He is a professor in electrical engineering department at Faculty of sciences & techniques, Fes, MOROCCO. His major research interests include Optical Components, Image, Audio and video processing. He is managing the optical and image processing research group.

**Ali AHAITOUF:** received the Ph.D. degrees in electronics from the Metz University in France 1992. He is a Professor in electrical engineering department at Faculty of sciences & techniques, Fes, MOROCCO, when he obtained the Doctor Title in Physics at 1998. His major research interests include Digital and Analog VLSI architecture, EMC Simulation and Physics of Semiconductor Components. He is managing the Microelectronics and Components research group.